\documentclass[acmsmall,screen]{acmart}

\usepackage[ruled,linesnumbered]{algorithm2e}
\usepackage{multirow}
\usepackage{listings}
\usepackage{booktabs}
\usepackage{multicol}
\usepackage{tabularx}
\usepackage{adjustbox}
\usepackage{subfig}
\usepackage{subcaption}
\usepackage{makecell}
\usepackage{pifont}
\usepackage{enumitem}
\usepackage{calligra}
\usepackage{mdframed}
\usepackage{fontawesome}
\usepackage{amsmath}
\usepackage{amsfonts}
\usepackage{graphicx}
\usepackage{textcomp}
\usepackage{xcolor}
\usepackage{balance}
\usepackage{caption}
\usepackage{wrapfig}
\usepackage{colortbl}
\usepackage{forest}
\usepackage{array}
\usepackage[most]{tcolorbox}

\usepackage{wrapfig}

\usepackage{booktabs}
\usepackage{colortbl}
\usepackage{xcolor}

\usepackage{forest}
\usepackage[most]{tcolorbox}

\def\BibTeX{{\rm B\kern-.05em{\sc i\kern-.025em b}\kern-.08em
    T\kern-.1667em\lower.7ex\hbox{E}\kern-.125emX}}

\lstdefinestyle{codeStyle}{
    basicstyle=\small\ttfamily,
    commentstyle=\color{green},
    keywordstyle=\color{blue},
    stringstyle=\color{blue},
    showstringspaces=false,
    breaklines=true,
    frame=lines,
    backgroundcolor=\color{white},
    captionpos=b,
    aboveskip=10pt,
    belowskip=10pt
}
\definecolor{lightgreen}{rgb}{0.9,1,0.9}
\definecolor{grayheader}{gray}{0.85}
\definecolor{codebackground}{RGB}{240,240,240} %

\newcommand{\appname}{{\sc IntentTester}\xspace}
\title{IntentTester: Intent-Driven Multi-agent Framework for Cross-Library Test Migration}

\setcopyright{cc}
\setcctype{by}
\acmDOI{10.1145/3797096}
\acmYear{2026}
\acmJournal{PACMSE}
\acmVolume{3}
\acmNumber{FSE}
\acmArticle{FSE068}
\acmMonth{7}
\acmSubmissionID{fse26maina-p1924-p}
\received{2025-09-03}
\received[accepted]{2025-12-22}

\begin{document}

\author{Yi Gao}
\orcid{0009-0000-2554-2381}
\affiliation{%
  \institution{Zhejiang University}
  \city{Hangzhou}
  \country{China}
}
\affiliation{%
  \institution{the Hangzhou High-Tech Zone (Binjiang) Institute of Blockchain and Data Security}
  \city{Hangzhou}
  \country{China}
}
\email{gaoyi01@zju.edu.cn}

\author{Ziyuan Zhang}
\orcid{0009-0001-2727-7806}
\affiliation{%
  \institution{Zhejiang University}
  \city{Hangzhou}
  \country{China}
}
\email{ziyuanzhang@zju.edu.cn}

\author{Xing Hu}
\orcid{0000-0003-0093-3292}
\affiliation{%
  \institution{Zhejiang University}
  \city{Hangzhou}
  \country{China}
}
\email{xinghu@zju.edu.cn}

\author{Xiaohu Yang}
\orcid{0000-0003-4111-4189}
\affiliation{%
  \institution{Zhejiang University}
  \city{Hangzhou}
  \country{China}
}
\email{yangxh@zju.edu.cn}

\author{Xin Xia}
\orcid{0000-0002-6302-3256}
\authornote{Corresponding Author}
\affiliation{%
  \institution{Zhejiang University}
  \city{Hangzhou}
  \country{China}
}
\affiliation{%
  \institution{the Hangzhou High-Tech Zone (Binjiang) Institute of Blockchain and Data Security}
  \city{Hangzhou}
  \country{China}
}
\email{xin.xia@acm.org}

\begin{abstract}
Unit tests capture both functional checks and domain-specific knowledge, but this knowledge remains locked within individual projects and is rarely reused across libraries with overlapping functionality.
Existing migration techniques based on structural code mappings (e.g., API signatures) often break down under divergent designs or cross-language settings, resulting in non-executable migrated tests. 
In this paper, we present \appname, a multi-agent framework for intent-driven test reuse. 
Instead of translating raw code, \appname abstracts tests into a language-agnostic Test Description Language (TDL), aligns them with semantically related entities and dependencies in a repository graph, and synthesizes executable tests through LLM-guided reasoning and iterative validation. 
This design enables cross-library and cross-language migration without manual intervention, producing migrated tests that existing structure-mapping approaches cannot achieve.
We evaluate \appname on nine open-source projects across three domains (JSON, HTML, and Time) and two languages (Java and Python). 
\appname generates 2,776 syntactically correct tests with 85\% correctness; in comparison, the two baselines achieve 51\% and 43\%.
Among them, 2,410 tests executed successfully, yielding a 74\% effectiveness rate.
Beyond higher success rates, \appname also surfaced previously unknown defects—including stack overflows, null dereferences, and parsing inconsistencies, several of which have been acknowledged or patched by maintainers.
Our results show that intent-driven migration shifts the focus from code mappings to semantic alignment, allowing practical cross-library and cross-language test reuse while improving test quality and exposing implementation flaws.
\end{abstract}


\maketitle

\section{Introduction}
Unit testing is the backbone of software reliability, and open-source communities such as GitHub now host millions of high-quality unit tests created by developers to validate a wide range of functionalities~\cite{yuan2024evaluating,aniche2021developers,wang2024software,ardic2025qualitative,khatami2023quality,schafer2024empirical,zhang2026automated}.
These tests not only verify functional correctness but also embody valuable domain knowledge.
However, this rich body of testing knowledge is locked inside individual projects, rarely reused across libraries that offer similar functionality~\cite{wang2020empirical,huang2022characterizing,schafer2024empirical,tang2024chatgpt}.
When multiple libraries implement the same functionality, developers end up writing tests from scratch even though equivalent tests already exist elsewhere.
For example, JSON parsing has both \textit{Gson} (Java) and \textit{SimpleJson} (Python), while HTML parsing has \textit{Jsoup} and \textit{JFiveparse} (both in Java).
In practice, these overlapping domains mean developers duplicate effort, slow down the adoption of new libraries, and risk leaving defects undiscovered when test suites are inconsistently reimplemented~\cite{gao2024mut}.

A natural solution is cross-library test migration, which aims to reuse tests across libraries with similar functionality.
However, current approaches such as \textsc{MUT}~\cite{gao2024mut} and \textsc{METALLICUS}~\cite{jha2023jtestmigbench} rely on structural code mappings, aligning API signatures or code patterns between libraries and then translating the test code accordingly.
This mapping-first paradigm suffers from two fundamental limitations:
First, \textit{structural and linguistic heterogeneity}. 
Libraries that provide similar features often expose them through divergent API designs, coding styles, or even across different programming languages, making structural mappings sparse or infeasible.
Second, \textit{manual adaptation overhead}. 
Even when partial mappings exist, state-of-the-art tools typically require developers to manually adjust the generated test before it can run, introducing extra costs and limiting automation.
As a result, most prior tools struggle to deliver runnable tests in diverse or cross-language settings, limiting their ability to fully leverage the potential of test knowledge reuse.
To enable cross-library test reuse and facilitate the sharing of best practices, we identify two core challenges:

\noindent \textbf{Challenge 1: Structural incompatibility}.
Even when libraries implement the same functionality, their internal code structures and API designs diverge significantly. 
Direct structural mappings are scarce, and differences in design philosophy, coding styles, or language constructs make API-level alignment unreliable.

\noindent \textbf{Challenge 2: Dependency completeness}.
Executable tests rarely depend on a single API alone. 
They require constructing fixtures, initializing parameters, and chaining return values to form valid execution paths. 
When these transitive dependencies are not captured, migrated tests fail at runtime due to missing initializations or parameter mismatches.

To overcome these limitations, we present \appname, a multi-agent framework for intent-driven test migration.
The key idea is to abstract tests based on their intent, concentrating on the functionality being validated, rather than on raw code or API structures.
\appname decomposes test migration into five collaborating agents: \ding{182} \textbf{Intent Abstractor Agent} converts a source test into a language-agnostic Test Description Language (TDL), capturing metadata, inputs, execution steps, and assertions.
\ding{183} \textbf{Intent Alignment Agent} maps each step in the TDL onto semantically related entities in the target repository graph, expanding dependencies to form a compact but complete context bundle.
\ding{184} \textbf{Planning Agent} evaluates whether the retrieved context is sufficient to complete the test as described in the TDL, rejecting any test that lacks the necessary dependencies.
\ding{185} \textbf{Test Migration Agent} constructs tests by synthesizing the TDL intent with the relevant context and reference patterns, ensuring alignment with the target repository's behavior.
\ding{186} \textbf{Verification Agent} validates the outputs of the previous agents, checking for consistency and sufficiency, and applies lightweight feedback when errors or omissions are detected.
This design shifts the problem from brittle code translation to semantic alignment and reasoning, enabling test intents to be migrated across libraries and across programming languages without manual intervention, thus broadening the scope of reusable tests.

\begin{figure*}
\centering
\includegraphics[width=\linewidth]{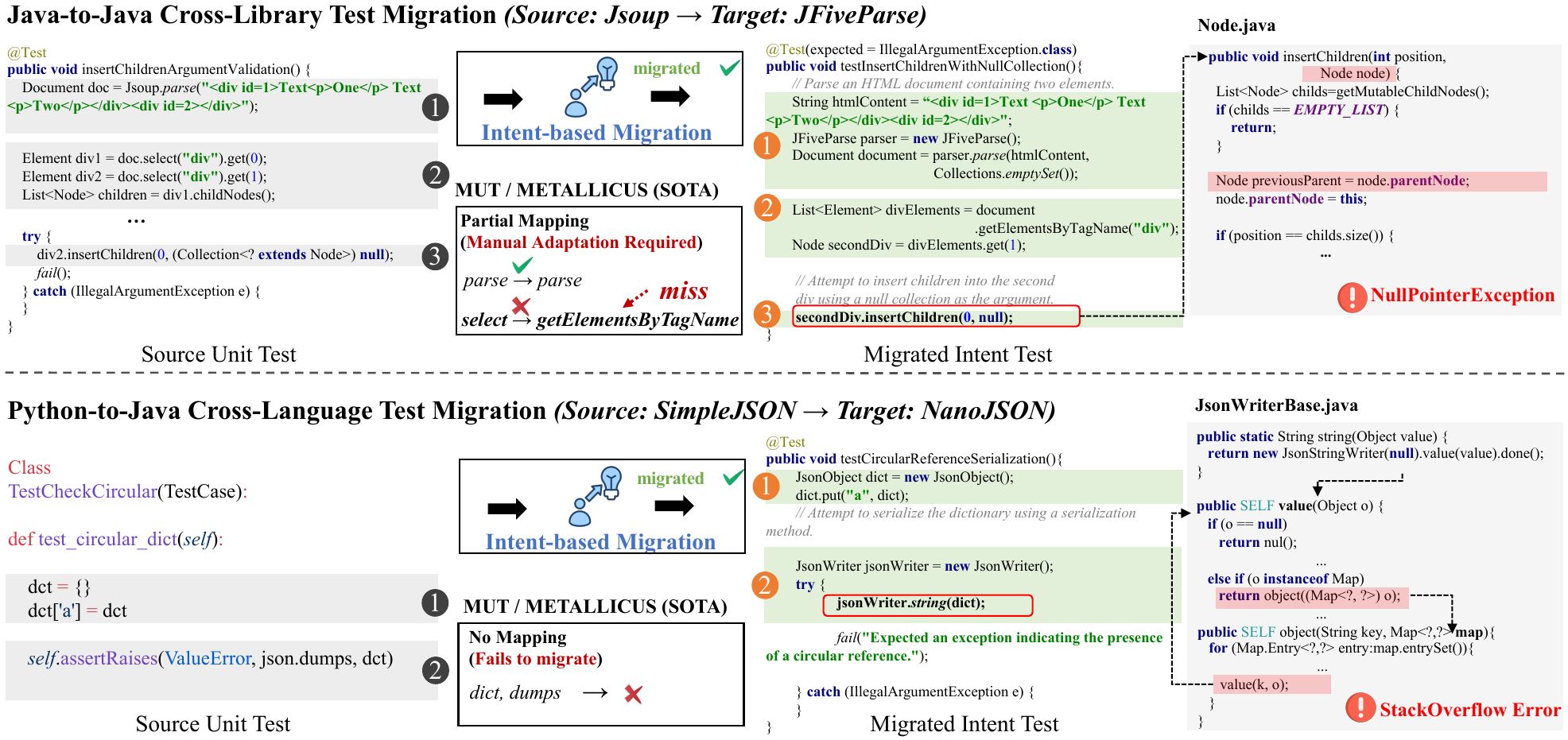}
\caption{Examples of cross-library and cross-language test migration. 
The HTML case (top) contrasts \textit{Jsoup} and \textit{JFiveParse} in handling null insertions, while the JSON case (bottom) shows a circular reference test migrated from \textit{SimpleJSON} to \textit{NanoJSON}. 
These cases demonstrate that intent-based migration preserves functionality and reveals hidden defects even when explicit structural mappings are incomplete or absent.}
\label{fig:moti}
\end{figure*}

We evaluate \appname on nine real-world open-source projects across three domains and two programming languages. 
From 2,058 source tests, we generate 5,536 sub-tests, of which 3,257 remain after filtering. 
Among these, \appname synthesizes 2,776 syntactically correct tests, achieving 85\% correctness—well above \textsc{MUT} (51\%) and \textsc{METALLICUS} (43\%).
Out of the correct tests, 2,410 execute successfully in the target repositories, corresponding to a 74\% effectiveness rate. 
Beyond execution success, \appname also uncovered 25 real defects across JSON, HTML, and Time libraries, including nested JSON parsing failures, missing HTML validation, and stack overflows in recursive serialization. 
Several of these issues have already been acknowledged or patched by maintainers.
These results confirm that intent-driven migration not only surpasses code-mapping baselines in producing runnable tests but also provides actionable value by strengthening test suites and revealing latent defects in widely used libraries.

The main contributions of this paper are as follows:
\begin{itemize}[leftmargin=*]
\item We identify the limitations of code-mapping approaches to test reuse and introduce intent-driven test migration as a new paradigm.
\item We present \appname, a multi-agent approach that leverages TDL abstraction, repository graph reasoning, and LLM-guided synthesis to enable cross-library test migration, with a replication package available at~\cite{intenttest}.
\item We build a fully automated pipeline supporting both Java and Python libraries, capable of generating runnable tests without manual adaptation. By decoupling migration from explicit mappings, it broadens reuse opportunities previously missed by prior approaches.
\item We conduct a large-scale study on nine repositories, showing that \appname improves syntactic correctness by 30–40\% and execution success by over 20\% compared to baselines, while uncovering 25 real defects.
\end{itemize}

\section{Motivation Example}
\label{sec:motivation}
Figure~\ref{fig:moti} presents two concrete intent-based test migration examples from real-world GitHub projects.
\textit{Jsoup} and \textit{JFiveParse} are both Java libraries designed for HTML parsing and manipulation.
In \textit{Jsoup}, the test reveals built-in argument validation that throws an \texttt{IllegalArgumentException} when attempting to insert \texttt{null} as a child node.
This design ensures that client projects using \textit{Jsoup} must handle the exception appropriately, thereby preventing unintended \texttt{null} insertions.

When this test is migrated to \textit{JFiveParse}, the same test intent is expressed through a newly migrated test. 
However, unlike \textit{Jsoup}, \textit{JFiveParse} only validates the parent node and omits checks on the inserted child nodes. 
As a result, executing \texttt{secondDiv.insertChildren(0, null)} dereferences a missing parent pointer, triggering a \texttt{NullPointerException}.
If this issue occurs in a client project relying on \textit{JFiveParse}, it could lead to an unexpected program crash.

This example highlights the limitations of state-of-the-art migration approaches such as \textsc{MUT} and \textsc{METALLICUS}, which rely on structural API mapping. 
While they can align basic calls like \texttt{parse} and \texttt{insertChildren}, they miss \texttt{select}, leading to \textbf{incomplete mappings} and requiring manual adaptation to make the migrated test executable.
However, by abstracting the source test into a language-agnostic intent, our approach achieves completeness in migration, automatically covering cases where structural mappings are partial or absent. 
This enables the generation of fully executable tests without manual intervention.

Another example involves JSON serialization libraries across different languages.
As shown in the lower-left part of Figure~\ref{fig:moti}, \textit{SimpleJSON} (Python) includes a dedicated test for handling circular references, where a dictionary key points to the dictionary itself. 
The test intent can be summarized as: detect circular references and raise an exception, ensuring that client projects cannot serialize self-referencing objects without explicit handling.
In contrast, \textit{NanoJSON} (Java) offers lightweight JSON serialization. 
Migrating the test intent to \textit{NanoJSON} produces a corresponding unit test that attempts to validate the same behavior. 
However, when executing the intent test in \textit{NanoJSON}, the project terminates with a \texttt{StackOverflowError}.
After further analysis, we identify that \textit{NanoJSON}’s \texttt{JsonWriterBase} processes \texttt{Map} objects by iterating over each key-value pair and recursively invoking the \texttt{value()} method on the values.
In the presence of a self-referencing object (e.g., \texttt{dict.put("a", dict)}), the serialization process enters an infinite recursion:
\texttt{value(dict)} → \texttt{object(map)} → \texttt{value("a", dict)} → \texttt{object(map)} → \texttt{value("a", dict)}, ultimately leading to a runtime crash.
This behavior represents a significant risk: if a client project using \textit{NanoJSON} inadvertently introduces a circular reference—or, more critically, if an attacker exploits this defect—the client crashes due to unhandled recursion.

As shown in Figure~\ref{fig:moti}, this case poses a critical challenge: there is \textbf{no structural mapping} between the \texttt{dumps} API and Java serialization entry points.
Unlike structure-based mapping, our approach aligns tests at the intent level, enabling the reuse of Python test intents in Java libraries even when APIs share no structural commonality, thereby surpassing the structural limitations of prior tools.
This design allows us to migrate tests across both libraries and languages, and in practice, it exposes hidden defects such as unhandled recursion in \textit{NanoJSON}.

\begin{figure}
\centering
\includegraphics[width=0.95\linewidth]{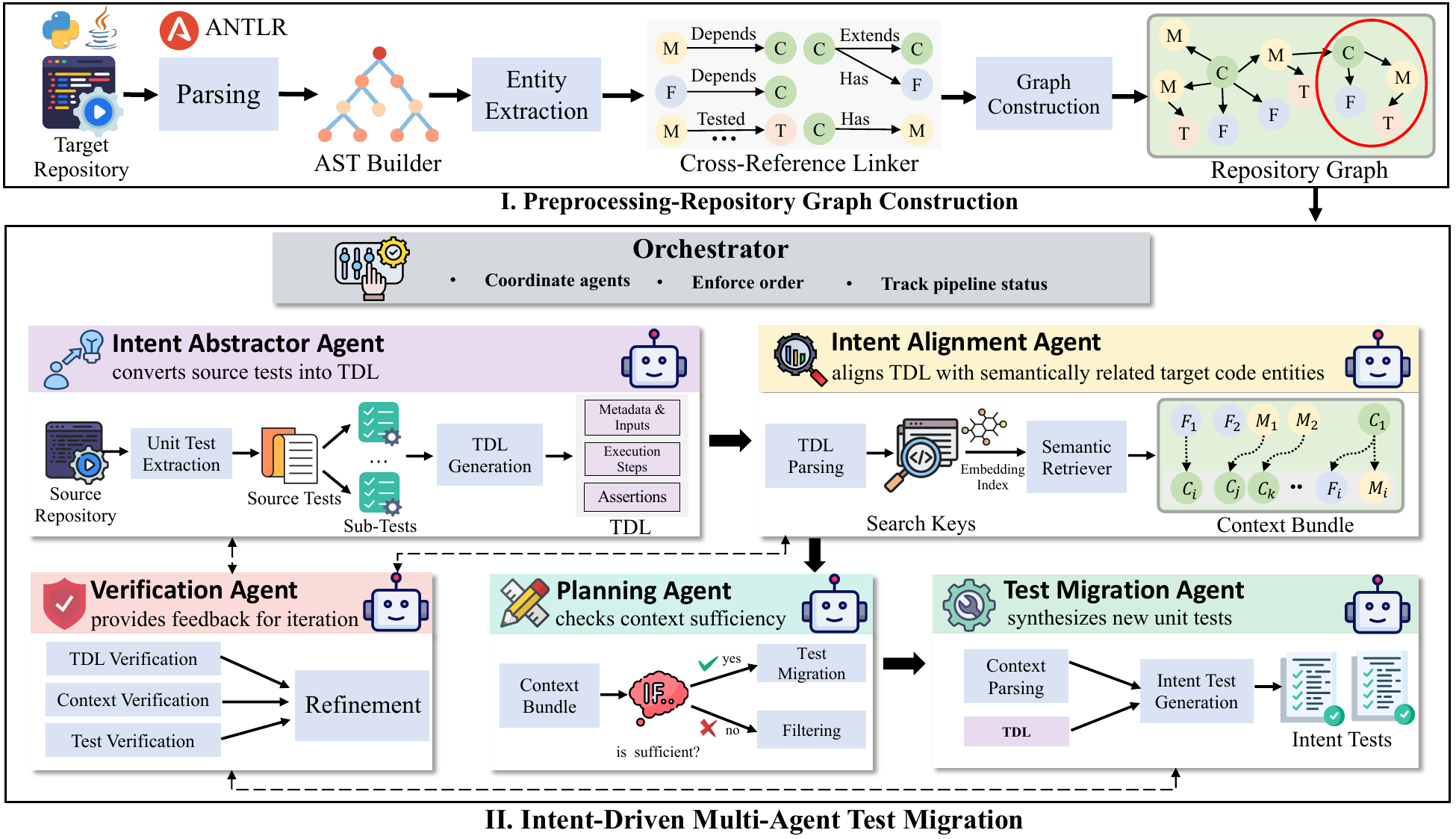}
\caption{Overall Pipeline Intent-Driven Test Migration.}
\label{fig:approach}
\end{figure}

We report the issues identified through intent-based testing to the respective library maintainers. 
Both \textit{NanoJSON} and \textit{JFiveParse} authors confirm the validity of our findings. 
Notably, the maintainers of \textit{JFiveParse} have already addressed the issue by adding null checks to their repository, effectively mitigating the NPE risk. 
\section{Approach}
\label{sec:approach}


Our approach consists of a preprocessing step (sec.~\ref{sec:preprocessing}) that builds a repository graph for the target repository, followed by a multi-agent framework \appname for intent-driven test migration.
The framework involves five agents: the Intent Abstractor (TDL conversion, sec.~\ref{sec:abstractorAgent}), Intent Alignment (semantic retrieval, sec.~\ref{sec:alignmentAgent}), Planning (context sufficiency, sec.~\ref{sec:planningAgent}), Test Migration (test synthesis, sec.~\ref{sec:migrationAgent}), and Verification (validation and feedback, sec.~\ref{sec:verificationAgent}).
An Orchestrator coordinates these agents to ensure a coherent and reliable workflow.

\begin{table}
\centering
\caption{Edge types in the repository graph, capturing structural and behavioral relations among classes, methods, fields, and tests.}
\footnotesize
\setlength{\tabcolsep}{3.8pt}
\renewcommand{\arraystretch}{1.06}
\begin{tabularx}{0.98\columnwidth}{l l X}
\toprule
\textbf{Edge Type} & \textbf{Scope} & \textbf{Description and Example} \\
\midrule
INHERITS           & Class $\leftrightarrow$ Class & Class extends a superclass (e.g., \texttt{JsonObject} $\to$ \texttt{JsonElement}). \\
IMPLEMENTS         & Class $\leftrightarrow$ Class & Class implements an interface (e.g., \texttt{Moment} $\to$ \texttt{UnixTime}). \\
\midrule
HAS\_METHOD        & Class $\leftrightarrow$ Method & Method declared in a class (e.g., \texttt{JsonParser.parseString}). \\
HAS\_FIELD         & Class $\leftrightarrow$ Field  & Field declared in a class (e.g., \texttt{JsonArray.elements}). \\
\midrule
CALLS\_METHOD      & Method $\leftrightarrow$ Method & One method calls another (e.g., \texttt{parseString} $\to$ \texttt{readToken}). \\
ACCESSES\_FIELD    & Method $\leftrightarrow$ Field  & Method reads/writes a field (e.g., \texttt{iterator} accesses \texttt{elements}). \\
\midrule
DEPENDS\_ON\_CLASS & Method/Field $\leftrightarrow$ Class & Parameter or field type depends on a class (e.g., \texttt{parse} depends on \texttt{JsonReader}). \\
RETURNS\_CLASS     & Method $\leftrightarrow$ Class & Method return type relation (e.g., \texttt{deepCopy} returns \texttt{JsonElement}). \\
\midrule
TESTS\_METHOD      & Test $\leftrightarrow$ Method & Unit test validates a method (e.g., \texttt{testParse} $\to$ \texttt{parse}). \\
USES\_CLASS        & Test $\leftrightarrow$ Class  & Test constructs/initializes a class (e.g., \texttt{new JsonArray()}). \\
ASSERTS\_FIELD     & Test $\leftrightarrow$ Field  & Test checks a field state (e.g., assert size of \texttt{elements}). \\
\bottomrule
\end{tabularx}
\label{tab:entity}
\vspace{-0.2cm}
\end{table}

\subsection{Preprocessing}
\label{sec:preprocessing}
In the preprocessing step, we extract code entities from the target repository to construct a corresponding repository graph that supports intent-based alignment. 
This step is performed once and can be subsequently reused for multiple intent-based test migrations.

\subsubsection{Code Entity Extraction}
The structure of a software repository and the relationships among its code entities are typically intricate~\cite{ma2024understand}: classes participate in inheritance hierarchies, methods invoke external modules, fields reference user-defined types, and unit tests often interact with multiple interconnected components. 
Such complexity makes naive parsing insufficient for supporting intent-based test migration, as incomplete or imprecise extraction can lead to missing dependencies and invalid migrated tests.

To address these challenges, we design a structured extraction pipeline that captures fine-grained entities and their structural dependencies.
Specifically, we extract four categories of entities—classes, fields, methods, and unit tests—and explicitly link them through inheritance, type references, method invocations, and test-to-method relations. 
This explicit modeling ensures that dependencies such as parameter types and indirect invocations are preserved, enabling downstream agents to reason over a complete and semantically consistent view of the repository.
Technically, we leverage \textsc{ANTLR}~\cite{antlr}, an extensible parser framework, to process source code and generate Abstract Syntax Trees (ASTs). 
We then traverse the ASTs to extract entity definitions and cross-entity references, enriching each entity with metadata such as type signatures, dependencies, and file locations.
The pipeline currently supports Java and Python, and can be extended to other ecosystems via \textsc{ANTLR}’s multi-language support.

\subsubsection{Graph building and relationship modeling}
After extracting code entities, we organize them into a repository graph that encodes their relationships, enabling downstream agents to trace multi-hop dependencies essential for intent-driven test migration.

\begin{center}
\begin{minipage}{0.9\columnwidth} 
\begin{algorithm}[H]\small
\caption{Agent Orchestration for Intent-Driven Test Migration}\label{algo:orchestration}
\KwIn{$sourceTest$: a test from source repo, $targetGraph$: repository graph of target repo}
\KwOut{$migratedTest$: executable test or $\varnothing$ if migration fails}

\BlankLine
\SetKwProg{Fn}{Function}{:}{}
\Fn{CoordinateAgents($sourceTest, targetGraph$)}{
    $tdl \gets$ IntentAbstractAgent.extract($sourceTest$)\;
    $context \gets$ IntentAlignmentAgent.retrieve($tdl, targetGraph$)\;

    \If{Planning.isSufficient($tdl, context$) = False}{
        $context \gets$ IntentAlignment.refine($tdl, targetGraph$)\;
        \If{Planning.isSufficient($tdl, context$) = False}{
            \Return $\varnothing$ 
        }
    }
    $candidateTest \gets$ TestMigrationAgent.synthesize($tdl, context$)\;
    $result \gets$ VerificationAgent.validate($candidateTest$)\;

    \If{$result$ = pass}{
        \Return $candidateTest$\;
    }
    \Else{
        $context \gets$ VerificationAgent.feedback($tdl, context$)\;
        \Return CoordinateAgents($sourceTest, targetGraph$)\;
    }
}
\end{algorithm}
\end{minipage}
\end{center}

We define the edge types summarized in Table~\ref{tab:entity}, which capture both structural and behavioral relationships among code entities. 
These include inheritance hierarchies, method invocations, type dependencies, field accesses, and test-to-method links obtained by statically resolving unit test calls to their target methods. 
By covering both structural (e.g., \textit{HAS\_METHOD}, \textit{DEPENDS\_ON\_CLASS}) and behavioral (e.g., \textit{CALLS\_METHOD}, \textit{TESTS\_METHOD}) links, the graph captures repository semantics comprehensively.

Modeling only call graphs or inheritance hierarchies omits important interactions necessary for test reuse, such as indirect dependencies across classes. 
To address this, we explicitly define a schema that integrates multiple relationship types (structural and behavioral) into a unified representation. 
This ensures that entities such as tests, methods, and fields are connected through all relevant dependencies, rather than isolated by a single view of the repository. 
For efficient storage and querying, we use \textit{Neo4j}~\cite{neo4j}, where entities are represented as labeled nodes and relationships as typed edges, enabling scalable retrieval of subgraphs relevant to a given test intent.

To further support intent-based alignment, each node is enriched with a concise textual description of its functionality. 
Since comments are often incomplete or outdated, we employ an LLM-based summarization strategy: given each entity’s signature and local context (e.g., method body or class definition), the model generates a short natural-language description (e.g., \textit{Creates a deep copy of this element and all its children} for \texttt{deepCopy}). 
These summaries serve as semantic annotations of structural entities and directly support the alignment of test intents with repository nodes (see Sec.~\ref{sec:alignmentAgent}).

\subsection{Intent-Driven Multi-Agent Test Migration}
\subsubsection{Orchestration and Workflow Control}
\label{sec:orchestration}
Cross-library test reuse faces challenges from language differences, library designs, and testing logic. 
To address these, we decompose the task into specialized agents coordinated by an Orchestrator. 
As shown in Algorithm~\ref{algo:orchestration}, the Orchestrator first invokes the \textit{Intent Abstractor Agent} to derive the TDL representation, then calls the \textit{Intent Alignment Agent} to retrieve related context, and the \textit{Planning Agent} to assess the sufficiency of the retrieved context. 
Finally, the \textit{Test Migration Agent} synthesizes executable tests, while the \textit{Verification Agent} validates the result and, if necessary, feeds corrections back into the pipeline.

\begin{figure*}
\centering
\includegraphics[width=0.98\linewidth]{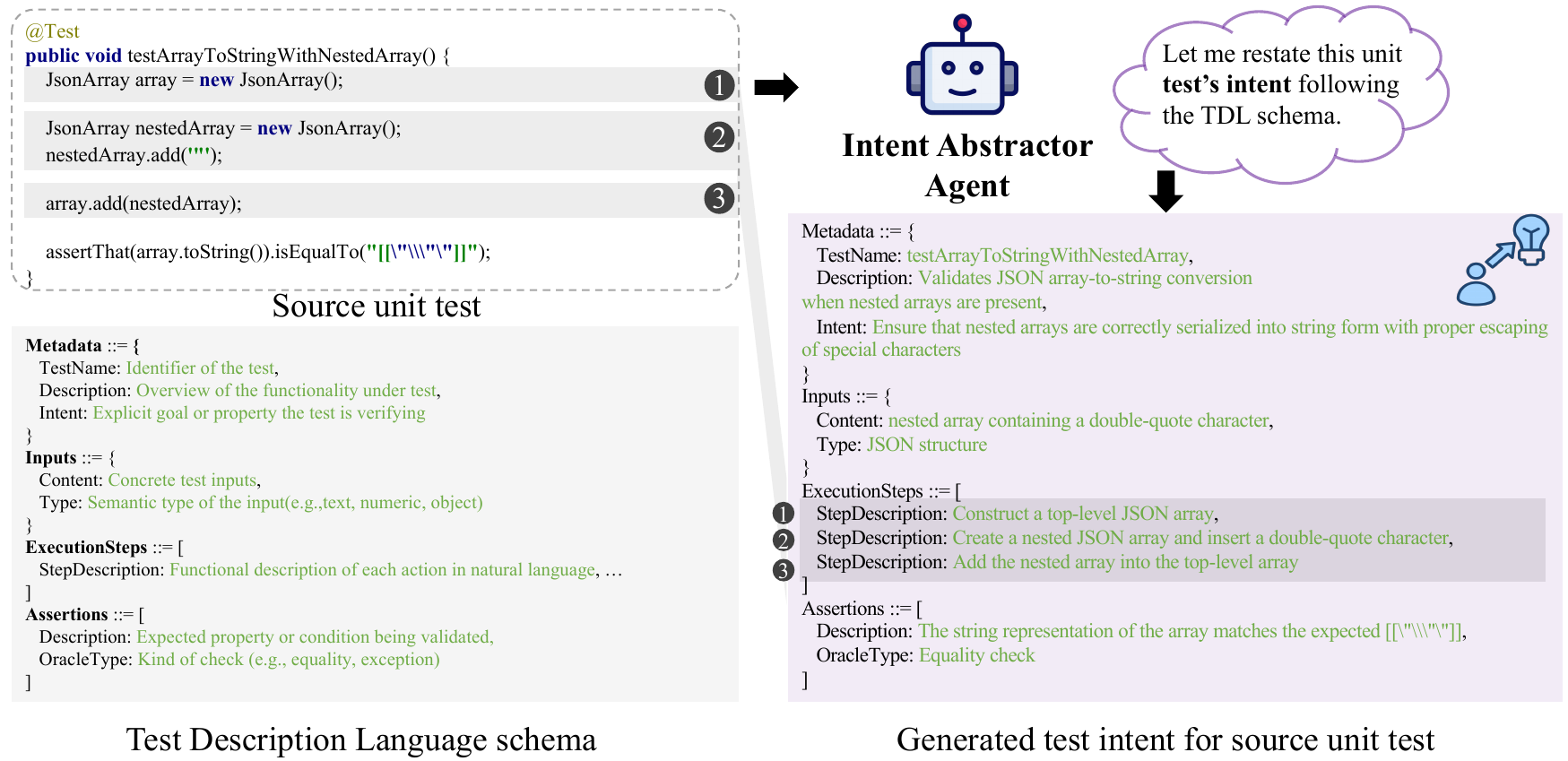}
\caption{Example of transforming a source unit test into its TDL representation, capturing metadata, inputs, execution steps, and assertions.}
\label{fig:tdl_exp}
\vspace{-0.4cm}
\end{figure*}

\subsection{Test Intent Abstraction}
\label{sec:abstractorAgent}
The \textit{Intent Abstractor Agent} transforms source tests into a unified representation by first decomposing complex tests into smaller units and then converting each unit into a TDL encoding its functional intent.

\subsubsection{Test Splitting}
Source tests are often multi-purpose, combining transferable logic (e.g., parsing) with library-specific mechanisms (e.g., error handling), which reduces reusability. 
To enable fine-grained reuse and improve the robustness of later agents, we decompose multi-purpose tests into simpler sub-tests, each aligned with a single functional intent.

Unlike prior work on test decomposition~\cite{gao2025automated}, which focused on analyzing test smells, our splitting procedure is redesigned for intent abstraction. 
First, splitting is tightly integrated with TDL generation: each sub-test is immediately converted into a TDL unit, guaranteeing direct usability for downstream agents. 
Second, the rules are extended to support both Java and Python tests, enabling cross-language consistency. 
Third, splitting improves fault isolation: if one sub-test is incorrectly migrated or fails during execution, the error is localized rather than invalidating the entire original test. 
Moreover, subsequent agents provide additional safeguards—Planning assesses the adequacy of retrieved context and Verification checks semantic correctness—so potential errors introduced during splitting do not propagate unchecked.

\subsubsection{TDL Generation}
The TDL captures the core functional intent of a test in a structured, language-agnostic format, enabling reuse across repositories that share functionality but differ in implementation details, programming language, or API design. 
Traditional tools such as \textsc{MUT} rely on structural mappings, which fail when similar behaviors are expressed through divergent code structures (Figure~\ref{fig:moti}). 
In contrast, TDL abstracts away implementation details and encodes only the essential test intent, allowing migration even in the absence of explicit mappings.

As shown in Figure~\ref{fig:tdl_exp}, a TDL consists of four components: test data, test setup, focal method, and assertions. 
Given a source test and the predefined schema, the \textit{Intent Abstractor Agent} transforms the test into its TDL using a prompt template that guides the model to restate each case according to the schema. 
In this example, the test that verifies the serialization of nested JSON arrays is represented as: (i) creating nested arrays, (ii) serializing them, and (iii) asserting correctness of the string output. 
Such representations are both descriptive and reusable, supporting intent-based retrieval from the target repository graph in subsequent steps. 
By uniformly applying TDL to all source tests, we decouple test intent from repository-specific implementation, thereby enabling effective cross-library and cross-language migration beyond the reach of structure-mapping approaches. 
Due to space limitations, we omit the full prompt design, but all templates are provided in our replication package for transparency and reproducibility~\cite{intenttest}.

\subsection{Semantic Context Alignment}
\label{sec:alignmentAgent}
Once the source test is abstracted into TDL, the challenge lies in identifying the target repository entities that can fulfill the described functionality. 
A single test intent often spans multiple classes, methods, and fields, and only complete dependency coverage enables executable test synthesis.

We highlight three key challenges: (i) each TDL description may correspond to multiple candidate entities under cross-language or stylistic variations; (ii) capturing only the most similar node is insufficient without its dependencies such as parameters, return values, and invocation chains; and (iii) large amounts of unrelated code reduce generation quality, making it essential to construct a minimal yet sufficient context for downstream test synthesis.

\begin{figure}
\centering
\includegraphics[width=\linewidth]{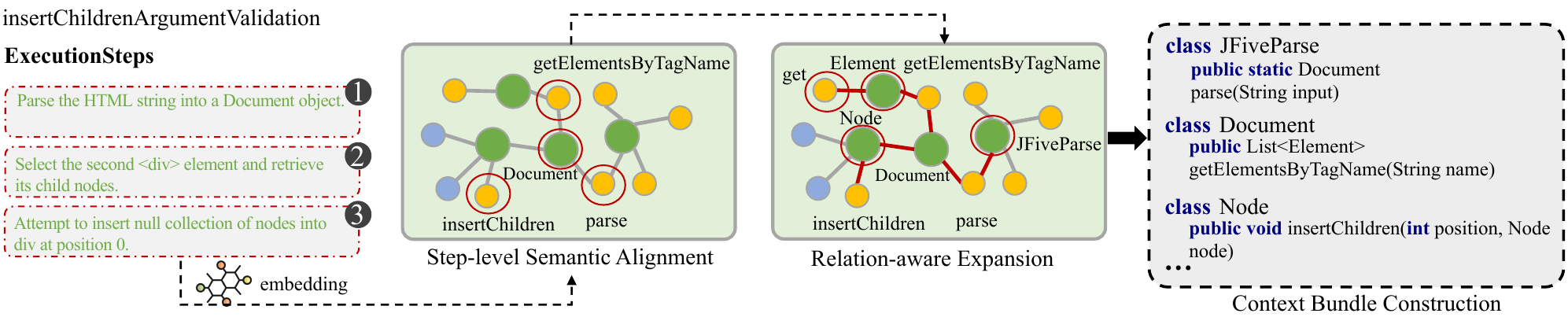}
\caption{Semantic alignment of TDL steps with repository graph entities, resulting in a context bundle.}
\label{fig:alignment}
\end{figure}

To address these challenges, the \textit{Intent Alignment Agent} decomposes the TDL into structured search units and performs a three-step retrieval-and-expansion process over the repository graph.
\textbf{(1) Step-level Semantic Querying}. Each \texttt{ExecutionStep} in the TDL is treated as a query unit. 
As shown in Figure~\ref{fig:alignment}, in the \texttt{insertChildrenArgumentValidation} case, a step such as \textit{Parse the HTML string into a Document} is encoded into embeddings and searched against the repository graph. 
We utilize \textit{MiniLM}~\cite{miniLM} as the embedding model because it offers a practical balance between accuracy and efficiency, enabling scalable retrieval across large repositories without sacrificing semantic precision. 
For each query, we perform a k-NN search  over the graph index to retrieve the top-k semantically similar nodes (implemented with \textit{FAISS}~\cite{faiss}). 
This design prioritizes recall, while subsequent agents (relation expansion and planning) filter and validate candidates to ensure executability. 
Retrieving multiple candidates instead of a single \textit{best match} directly addresses the challenge of cross-language and stylistic variations, where no exact signature correspondence exists. 
Unlike prior approaches based on explicit API matching, this step enables broader semantic coverage while preserving retrieval efficiency.
\textbf{(2) Relation-aware Expansion}. The retrieved nodes are then expanded along typed edges in the repository graph (e.g., \textit{INVOKES}, \textit{HAS\_PARAM}, \textit{RETURNS}). 
For example, retrieving \texttt{insertChildren} also pulls in its parameter class \texttt{Node} and the constructor path required to instantiate it, ensuring that the generated test satisfies all required dependencies for execution.
\textbf{(3) Context Bundle Construction}. Finally, the relevant entities are assembled into a \textit{Context Bundle}—a compact subgraph containing the minimal set of classes, methods, and dependencies required to fulfill the test intent. 
As shown in Figure~\ref{fig:alignment}, this bundle includes \texttt{JFiveParse.parse}, \texttt{Document.getElementsByTagName}, and \texttt{Node.insertChildren}, and serves as the input for subsequent test generation.

Compared with existing approaches such as \textsc{MUT} and \textsc{METALLICUS} that rely on explicit API mappings, \textit{Intent Alignment Agent} supports multi-to-multi, cross-language semantic alignment and ensures dependency completeness via relation-aware expansion. 
This design allows the \appname to bridge structural gaps where no API correspondence exists, enabling test migration in cases that prior approaches cannot handle.

\subsection{Context-Aware Planning}
\label{sec:planningAgent}
The \textit{Planning Agent} determines whether the retrieved context is sufficient for constructing an executable unit test. 
In typical cases, the context bundle already covers the entities required to instantiate inputs and invoke the target functionality. 
However, certain projects exhibit more intricate construction patterns. 
For example, in \textit{NanoJSON}, initializing a JSON parser requires first constructing a \texttt{JsonTokener}, which itself depends on a \texttt{StringReader}; only with these chained dependencies resolved can the parser be executed correctly.

To address such cases, the agent performs a lightweight validation-and-expansion process. 
It first checks whether all entities required by the TDL intent (e.g., constructors, invoked methods, expected assertions) are included in the context bundle. 
If some dependencies are missing, the agent performs a one-hop expansion along typed edges in the repository graph (e.g., \texttt{HAS\_PARAM}, \texttt{RETURNS}, \texttt{INVOKES}). 
If coverage remains incomplete after expansion, the intent is deemed unsupported in the target library and safely discarded. 
This mechanism not only avoids spurious generation when the functionality does not exist in the target repository, but also ensures that test generation proceeds only when the context is both sufficient and compact, tolerating minor retrieval gaps while preventing false migrations.

\subsection{Intent-Guided Test Synthesis}
\label{sec:migrationAgent}
\begin{figure}
\centering
\includegraphics[width=0.9\linewidth]{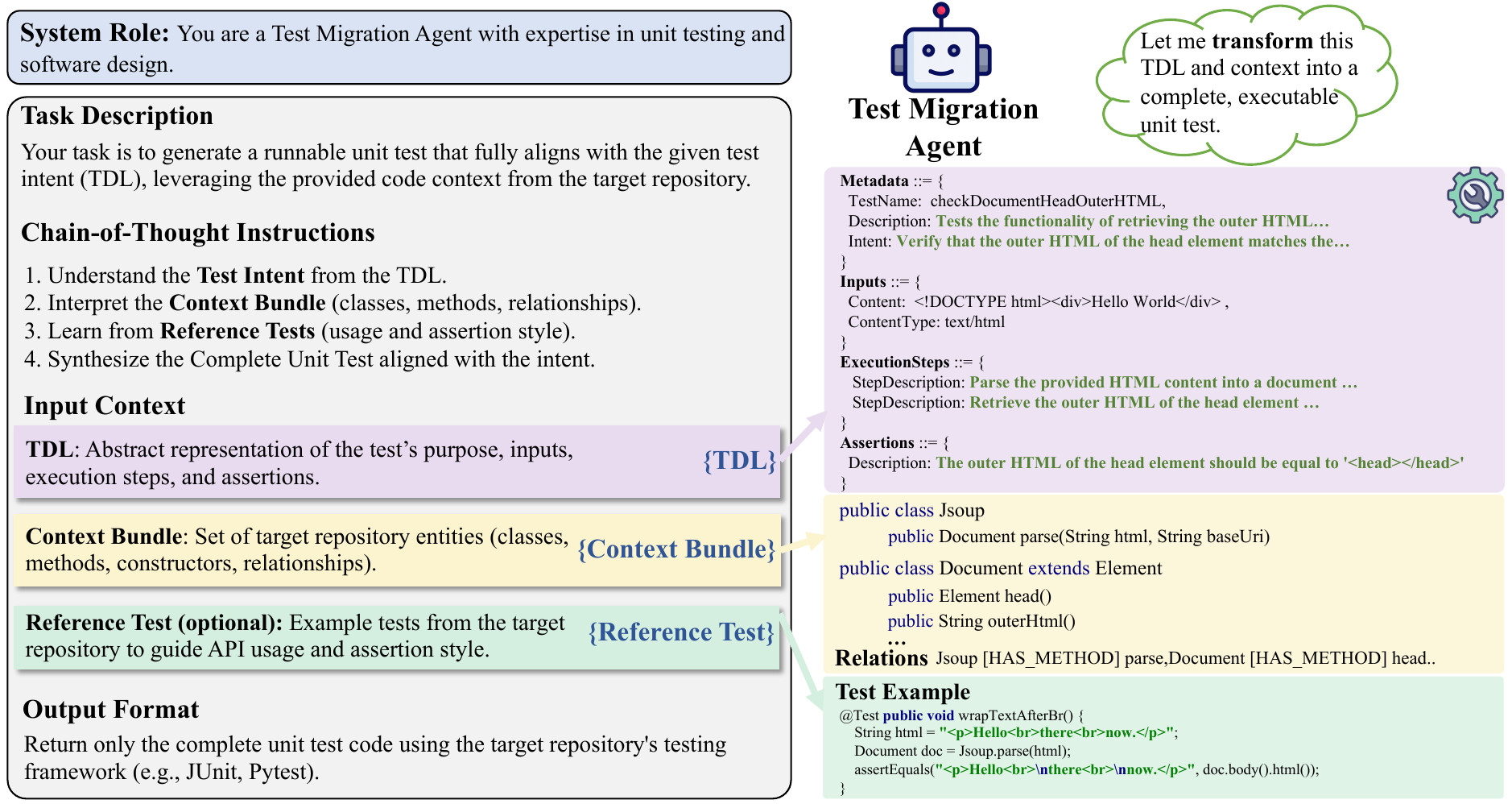}
\caption{Prompt Design of the Test Migration Agent, which integrates the abstracted test intent (TDL), the retrieved context bundle, and reference tests, and applies chain-of-thought prompting to synthesize complete executable unit tests in the target repository.}
\label{fig:prompt_template}
\end{figure}

The \textit{Test Migration Agent} is responsible for constructing complete unit tests from the extracted intent and retrieved context. 
To enable step-wise reasoning, we design a Chain-of-Thought (CoT) prompting paradigm~\cite{li2025structured,hou2024large}, which explicitly decomposes the generation process into a sequence of reasoning steps.
This CoT design mirrors the reasoning process of human testers: first clarifying the testing objective, then identifying the necessary entities and their relationships, and finally constructing executable code. 
As shown in Figure~\ref{fig:prompt_template}, the agent is guided through four steps:
\ding{182} \textbf{Understanding the test intent} as described in the TDL.
\ding{183} \textbf{Interpreting the Context Bundle}, including relevant classes, methods, and their relationships in the target repository.
\ding{184} \textbf{Learning from reference tests} retrieved in prior steps to capture realistic API usage and assertion styles.
\ding{185} \textbf{Synthesizing the complete test case}, ensuring alignment with both the functional intent and the structural constraints of the repository.

The core novelty lies in integrating functional intent (from TDL) with repository-specific usage patterns (from the Context Bundle and reference tests), enabling the \appname to migrate tests that are both semantically faithful and structurally valid.
Unlike prior approaches such as \textsc{MUT} or \textsc{METALLICUS}, which depend on rigid code mappings, handcrafted templates, or manual adjustments, our agent supports broader cross-library migration in a fully automated manner. 
This paradigm enables the reuse of tests even when explicit structural mappings are absent, thereby advancing automated test migration toward more realistic and heterogeneous software ecosystems.

\subsection{Iterative Validation and Feedback}
\label{sec:verificationAgent}
The outputs of individual agents are not always reliable in isolation: a generated TDL may omit essential details, the retrieved context may lack key dependencies, or the synthesized test may contain incorrect assertions. 
To address this, \appname integrates an iterative validation loop. 
Each intermediate artifact—TDL, context bundle, and migrated test—is checked for structural and semantic consistency. 
When inconsistencies are detected, the framework automatically feeds back error signals to the corresponding agent and triggers regeneration with a refined prompt. 
If errors persist after three iterations, the migration instance is discarded to prevent infinite retries, ensuring that the framework remains efficient and does not produce unreliable outputs.

This lightweight feedback loop ensures that incomplete or incorrect intermediate results do not propagate into the final output. 
More importantly, it reflects the practical reality of test construction: assembling an executable test often requires multiple refinements, especially under heterogeneous libraries. 
By embedding this iterative mechanism, \appname improves robustness without requiring human intervention, striking a balance between automation and reliability.

\section{Evaluation}
\label{sec:evaluation}
Our experiments are designed to address the following research questions:

\begin{itemize}[leftmargin=*]
\item \textbf{RQ1: What is the quality of the intent test code migrated by \appname?}
\item \textbf{RQ2: How effective are the intent tests migrated by \appname?}
\item \textbf{RQ3: How effective is TDL in generating intent tests?}
\item \textbf{RQ4: How does the absence of each agent affect migration success in \appname?}
\end{itemize}

\begin{table}[t]
\centering
\caption{Dataset overview. For each repository, \#Src = original tests in that repo; \#Intent$_\text{in}$ = tests \emph{migrated into} this repo from the other two repositories in the same domain (after splitting/filtering).}
\setlength{\tabcolsep}{3.8pt}
\renewcommand{\arraystretch}{1.06}
\begin{tabularx}{0.93\columnwidth}{l X l r r}
\toprule
\textbf{Domain} & \textbf{Repository} & \textbf{Language} & \textbf{\#Src} & \textbf{\#Intent$_\text{in}$}  \\
\midrule
\multirow{3}{*}{JSON}
 & Gson~\cite{gson}        & Java &  87 & 160 (= NanoJSON 82 + SimpleJSON 78) \\
 & NanoJSON~\cite{nanojson}    & Java & 82 & 165 (= Gson 87 + SimpleJSON 78)      \\
 & SimpleJSON~\cite{simplejson}   & Python & 78 & 169 (= Gson 87 + NanoJSON 82)      \\
\midrule
\multirow{3}{*}{HTML}
 & Jsoup~\cite{jsoup}       & Java & 283 & 203 (= JFiveParse 65 + Domonic 138)  \\
 & JFiveParse~\cite{jfiveparse}  & Java & 65 & 421 (= Jsoup 283 + Domonic 138)       \\
 & Domonic~\cite{domonic}       & Python & 138 & 348 (= Jsoup 283 + JFiveParse 65)  \\

\midrule
\multirow{3}{*}{Time}
 & Time4j~\cite{time4j}      &  Java & 92 & 204 (= Threeten 71 + Maya 133)       \\
 & Threeten~\cite{threeten}    &  Java & 71 & 225 (= Time4j 92 + Maya 133)         \\
 & Maya~\cite{maya}          & Python & 133 & 163 (= Time4j 92 + Threeten 71)   \\
\midrule
\textbf{Total} & \textbf{9 repositories} & & \textbf{1,029} & \textbf{2,058} \\
\bottomrule
\end{tabularx}
\label{tab:dataset}
\end{table}

\subsection{Experimental Setup}
\noindent \textbf{Dataset.} 
We collect nine widely used open-source libraries from GitHub, covering three domains: JSON processing, HTML parsing, and time manipulation, with three repositories per domain (six in Java and three in Python). 
This design enables evaluation in both intra-language and cross-language migration scenarios. 
To ensure quality and reproducibility, we require that each repository (i) compiles and runs correctly, (ii) has active commits within the last six months, and (iii) contains a sufficient number of valid test cases. 
We manually review and execute all tests, discarding failing or invalid ones. 
In total, we obtain 1,029 source tests across the nine repositories. 
From these, we construct 2,058 intent tests by deriving, for each repository, intent tests from the original cases of the other two repositories in the same domain. 
Table~\ref{tab:dataset} summarizes the domains, repositories, and the distribution of source and intent tests.

\noindent \textbf{Baseline.}
\textsc{METALLICUS}~\cite{sondhi2021mining} is a semi-automated framework for Java and Python that migrates tests by mining API-level correspondences. 
It extracts method signatures and documentation from the source tests, retrieves functionally similar candidates in the target repository, and adapts them using manually predefined templates. 
While effective in simple settings, this process depends on manual intervention when mappings are sparse or dependencies are complex. 
For fair comparison, we preserve its \emph{core mapping and retrieval components}, which represents the main technical contribution of \textsc{METALLICUS}.
To eliminate manual bias and ensure comparability with our approach, we replace its template-based adaptation with the same LLM (Llama-3.3-70B, default configuration) used in \appname for final code generation.
\textsc{MUT}~\cite{gao2024mut} targets Java–C++ test migration through class- and API-level mappings. 
It constructs correspondences via signature similarity, applies hard-coded translation rules, and finally requires manual adjustments to ensure executability. 
To ensure comparability, we retain its mapping stage but replace the rule-based adaptation with the same LLM used in \appname, enabling automatic completion and refinement of migrated tests.

\subsection{RQ1: What is the quality of the intent test code migrated by \appname?}
To evaluate the quality of migrated tests, we examine their syntactic correctness after integrating them into the target repositories. 
This metric reflects whether the produced test code compiles (Java) or runs without syntax errors and unresolved references (Python), serving as the foundation for subsequent execution-based evaluation (RQ2).
As shown in Table~\ref{tab:rq1}, \appname produces 2,776 syntactically correct tests out of 3,257, achieving an overall correctness rate of 85\%. 
In contrast, \textsc{MUT} and \textsc{METALLICUS} achieve only 51\% and 43\%, respectively. 
\appname achieves superior test syntax correctness through intent migration compared to \textsc{MUT} and \textsc{METALLICUS}, demonstrating the robustness of its intent-driven design.

This advantage is attributed to three core design principles. 
First, the TDL abstracts test functionality from syntactic and library-specific complexities, allowing tests to be re-expressed in a portable, intent-level form. 
Second, the \textit{Intent Alignment Agent} retrieves not only semantically relevant methods but also their dependent entities (e.g., parameters, constructors), ensuring dependency completeness within the Context Bundle. 
Third, the \textit{Planning} and \textit{Verification} agents filter out test intents unsupported by the target repository, preventing spurious test generation and ensuring that only valid contexts are migrated. 
Together, these agents jointly guarantee robustness, explaining why \appname consistently outperforms code-mapping-based baselines.

\begin{table*}[t]
\centering
\caption{Syntactic correctness of migrated intent tests (filter.\,$=$ tests after filtering, cor.\,$=$ syntactically correct tests, rate\,$=$ correctness rate).}
\resizebox{0.9\textwidth}{!}{
\begin{tabular}{llr|rrr|rr|rr}
\toprule
\multirow{2}{*}{\textbf{Domain}} & \multirow{2}{*}{\textbf{Repository}} & \multirow{2}{*}{\textbf{\#Tests}} 
& \multicolumn{3}{c|}{\textbf{\appname}} 
& \multicolumn{2}{c|}{\textbf{MUT}} 
& \multicolumn{2}{c}{\textbf{METALLICUS}} \\
\cmidrule(lr){4-6} \cmidrule(lr){7-8} \cmidrule(lr){9-10}
& & & filter. & cor. & rate & cor. & rate & cor. & rate \\
\midrule
\rowcolor{gray!10}
\multirow{3}{*}{JSON} 
 & Gson       & 534 & 416 & 367 & 88\% & 295 & 55\% & 225 & 42\% \\
 & NanoJSON   & 624 & 340 & 252 & 74\% & 190 & 50\% &  89 & 26\% \\
 & SimpleJSON & 685 & 650 & 634 & 97\% & 342 & 50\% & 311 & 48\% \\
\midrule
\rowcolor{gray!10}
\multirow{3}{*}{HTML} 
 & Jsoup      & 436 & 242 & 217 & 90\% & 160 & 54\% & 154 & 63\% \\
 & JFiveParse & 991 & 398 & 288 & 72\% & 230 & 49\% & 167 & 42\% \\
 & Domonic    & 959 & 311 & 247 & 79\% & 165 & 53\% & 106 & 34\% \\
\midrule
\rowcolor{gray!10}
\multirow{3}{*}{Time} 
 & Time4j     & 412 & 222 & 168 & 76\% & 120 & 52\% &  98 & 44\% \\
 & Threeten   & 411 & 338 & 326 & 96\% & 200 & 49\% & 124 & 37\% \\
 & Maya       & 484 & 340 & 277 & 82\% & 178 & 52\% & 135 & 40\% \\
\midrule
\textbf{Total} & & \textbf{5,536} & \textbf{3,257} & \textbf{2,776} & \textbf{85\%} & \textbf{1,880} & \textbf{51\%} & \textbf{1,409} & \textbf{43\%} \\
\bottomrule
\end{tabular}}
\label{tab:rq1}
\end{table*}

One illustrative case comes from migrating a \textit{Gson} test that verifies the serialization of a top-level integer into JSON. 
In \textit{Gson}, this intent is expressed through the use of \texttt{JsonWriter} together with a \texttt{StringWriter}, which requires an explicit \texttt{close()} operation to finalize the output. 
In contrast, the equivalent behavior in Python’s \textit{SimpleJSON} relies on \texttt{json.dump} combined with \texttt{StringIO}, where the output must be rewound using \texttt{seek(0)} before reading.
Although both tests validate the same intent: \textit{serialize integer to JSON and check that the result is 123}, the APIs differ fundamentally. 
No direct mapping exists between \texttt{JsonWriter.close()} and \texttt{StringIO.seek(0)}, nor between \texttt{writer.value()} and \texttt{json.dump}. 
Consequently, code-mapping-based tools such as \textsc{MUT} or \textsc{METALLICUS} cannot establish a usable correspondence.
However, \appname abstracts the test intent into a TDL representation and semantically aligns it with the target repository. 
By reasoning about functional equivalence rather than API signatures, it successfully generates a valid Python test that compiles and runs correctly. 
This example highlights how intent-driven abstraction allows cross-language and cross-library migration even in the absence of structural mappings.

Although correctness is substantially improved, \appname still encounters errors in a subset of tests. 
The dominant failure mode arises from complex parameter construction, where the target API requires unusually deep initialization chains—often violating best-practice design principles.
For example, in \textit{Time4j}, the method \texttt{MultiFormatParser.parse()} expects a \texttt{ChronoFormatter} object that depends on seven layers of nested reference types. 
Similarly, in \textit{JFiveParse}, constructing a \texttt{NodeMatchers} instance requires traversing a deeply coupled dependency graph. 
In such cases, when the retrieved Context Bundle does not capture all transitive dependencies, the migrated test fails compilation due to unresolved constructors. 
These failures are rare and concentrated in repositories with atypically deep constructor graphs.
This RQ shows that intent-driven test reuse achieves much higher syntactic correctness than code-mapping baselines. 
By leveraging TDL abstraction and Context Bundle reasoning, \appname consistently generates compilable tests across heterogeneous repositories.

\subsection{RQ2: How effective are the intent tests migrated by \appname?}
\subsubsection{Overall Effectiveness}
We evaluate the effectiveness of intent-based test migration by executing all syntactically correct tests migrated in RQ1 across nine repositories. 
As shown in Table~\ref{tab:rq2_overall}, \appname achieves an overall 72\% pass rate, with consistent performance across domains: JSON (73\%), HTML (71\%), and Time (71\%).
To further measure how effectively the migrated tests preserve intent rather than just compiling, we define \text{Effective Accuracy} as: $\emph{Effective Accuracy} = \frac{\text{Pass Tests}}{\text{Filtered Tests}}$, where \text{Filtered Tests} are those that passed the syntactic check in RQ1. 
This metric reflects the fraction of intent tests that both compile and run correctly, thereby preserving their intended functionality. 
By this measure, \appname attains 82\% in JSON and 68\% in both HTML and Time, confirming its ability to generate runnable and meaningful cross-repository tests.

Unlike baselines such as \textsc{MUT} and \textsc{METALLICUS}, which report only 51\% and 43\% syntactic correctness in RQ1, \appname advances further to provide end-to-end executable tests. 
Baselines often require additional manual adaptation before execution (e.g., re-specifying parameters, re-writing missing oracles, or handling untranslated original APIs in tests due to missing mappings), which limits their automation potential.

\begin{table}[t]
\centering
\caption{Execution results of intent tests in target repositories. Pass and Fail counts are reported with percentages; failures are categorized into F1–F4.}
\begin{tabular}{c|c|c|c|c|c|c|c}
\toprule
\multirow{2}{*}{\textbf{Domain}} & \multirow{2}{*}{\textbf{Pass}} & \multirow{2}{*}{\textbf{Fail}} & \multicolumn{4}{c|}{\textbf{Failure Categories}} & \multirow{2}{*}{\textbf{Accuracy}} \\
\cline{4-7}
 & & & F1 & F2 & F3 & F4 &  \\
\hline
JSON & 916 (73\%) & 337 (27\%) & 105 & 198 & 15 & 19 & 82\% \\
HTML & 531 (71\%) & 221 (29\%) & 106 & 97  & 8  & 10 & 68\% \\
Time & 549 (71\%) & 222 (29\%) & 155 & 52  & 2  & 13 & 68\% \\
\hline
\textbf{Total} & \textbf{1,996 (72\%)} & \textbf{780 (28\%)} & \textbf{366} & \textbf{347} & \textbf{25} & \textbf{42} & \textbf{74\%} \\
\bottomrule
\end{tabular}
\label{tab:rq2_overall}
\end{table}

\subsubsection{Failure Categorization}
We next analyze the failing cases to understand their root causes. 
Execution outcomes are divided into \textbf{Pass} (successful execution) and \textbf{Fail} (execution failure). 
For the latter, we conduct a fine-grained classification into four categories:
\ding{183} \textbf{F1 Invalid Adaptation} – failures caused by misuse of APIs or parameter mismatches despite syntactic correctness.
\ding{184} \textbf{F2 Design Differences} – cross-library divergences in semantics or API behavior, e.g., normalization policies or indexing conventions.
\ding{185} \textbf{F3 Defect Discovery} – actual defects uncovered in the target repository, such as null dereference or infinite recursion.
\ding{186} \textbf{F4 Feature Gaps} – functionality supported in the source repository but absent in the target, though potentially implementable.

As summarized in Table~\ref{tab:rq2_overall}, F1 and F2 account for most failures (over 90\%), reflecting expected challenges of adaptation and heterogeneous library design rather than flaws in \appname. 
Importantly, 25 failures correspond to real defects (F3), several of which have been acknowledged and fixed by maintainers, while 42 cases fall into F4, highlighting opportunities for feature enhancement.

\begin{figure*}
\centering
\includegraphics[width=0.95\linewidth]{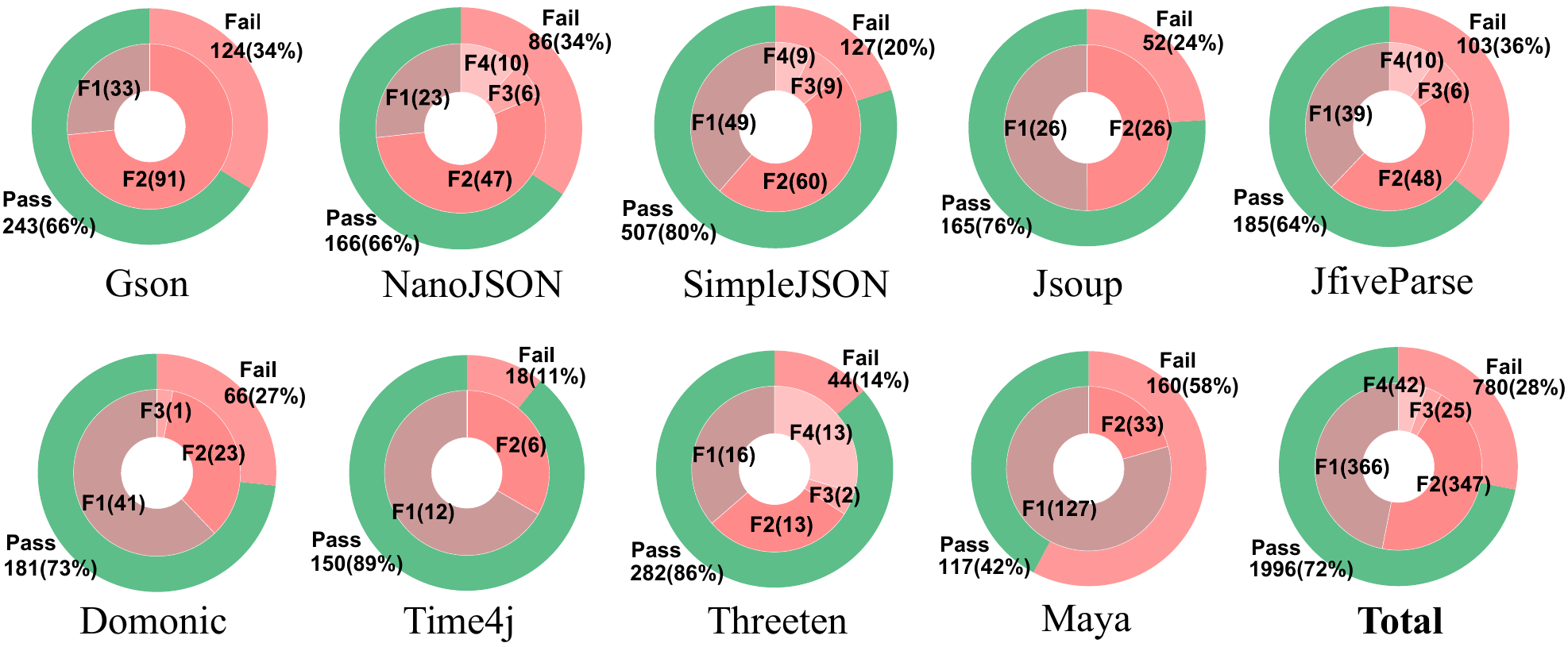}
\caption{Distribution of execution outcomes for migrated intent tests across different repositories, showing the proportion of successful and failed runs.}
\label{fig:passfail}
\end{figure*}

Figure~\ref{fig:passfail} presents the distribution of execution outcomes across all nine repositories. 
The JSON domain achieves the highest effective test rate at 82\%, while the HTML and Time domains both reach 68\%. 
This indicates that \appname reliably reuses tests across domains with distinct design styles.
Most failures fall into F1 Invalid Adaptation (47\%) and F2 Design Differences (45\%). 
Importantly, these are expected outcomes in cross-library migration rather than intrinsic limitations of \appname. 
By contrast, F3 Defect Discovery and F4 Feature Gaps collectively account for 9\% of failures but contribute the most value, exposing actual bugs and surfacing missing functionality.

\subsubsection{Case Study and Defect Categorization}
\textit{F1 Invalid Adaptation}. In \textit{JFiveParse}, an intent test fails when attempting to retrieve the root DOM element. 
The migrated test uses \texttt{getFirstElementChild}, which syntactically compiles but semantically retrieves the wrong entity. 
This results in an execution failure even though the test is valid at the code level. 
Such cases highlight that semantic API discrepancies remain a key challenge in cross-library test migration, but they also show where \appname could improve by enriching its context model with usage constraints.

\textit{F2 Design Differences}. Intent tests also expose subtle divergences in functionality. 
For example, \textit{Jsoup} automatically normalizes HTML tags (e.g., \texttt{<DIV>} → \texttt{<div>}), while \textit{JFiveParse} preserves casing. 
Similarly, in \textit{Domonic}, calling \texttt{get(-1)} returns the last element, whereas \textit{Jsoup} enforces strict index validation. 
These are not bugs but provide actionable insights for developers migrating or co-maintaining code across libraries.

Beyond adaptation challenges, a subset of execution failures revealed critical findings. 
As summarized in Table~\ref{tab:defects}, \appname uncovered 25 genuine defects across JSON, HTML, and Time repositories, including incorrect handling of deeply nested JSON, missing validation of HTML nodes, unhandled \texttt{null} values, and stack overflows in recursive serialization. 
We reported these issues to maintainers, of which six have been acknowledged and three already patched. 
In contrast, baselines such as \textsc{MUT} and \textsc{METALLICUS} could not uncover such defects, since their migrated tests failed to execute reliably.
In addition, 42 cases fall under feature gaps, where tests exercised functionality absent from the target library but potentially useful (e.g., automatic HTML normalization in \textit{JFiveParse}). 
Unlike baselines, which fail to reach execution due to missing mappings or incomplete contexts, \appname produces runnable tests that expose semantic mismatches across libraries. 
Overall, F1–F2 represent adaptation and heterogeneity challenges, while F3–F4 highlight the added value of \appname: identifying real defects and surfacing enhancement opportunities. 
These findings demonstrate that the value of \appname extends beyond raw pass rates: intent-driven tests enable cross-library semantic validation, uncover real defects, and surface meaningful opportunities for enhancement.

\begin{table}[t]
\centering
\caption{Classification of real defects uncovered by \appname. Each category includes representative error types.}
\begin{tabularx}{0.97\linewidth}{l X c}
\toprule
\textbf{Category} & \textbf{Representative Error Types} & \textbf{Count} \\
\midrule
HTML/XML Parsing     & Missing structure validation; incorrect DOM node handling & 4 \\
JSON Parsing         & Unicode misparsing; failure on deeply nested JSON; precision loss & 7 \\
Date/Time Handling   & Incorrect format conversion; date calculation mismatch & 2 \\
Data Validation      & Unhandled \texttt{null}; generation of invalid HTML tags & 3 \\
Exception Handling   & Null pointer dereference; infinite recursion $\rightarrow$ stack overflow & 6 \\
Format Conversion    & BigInt mis-conversion; wrong type casting & 3 \\
\midrule
\textbf{Total}       & & \textbf{25} \\
\bottomrule
\end{tabularx}
\label{tab:defects}
\vspace{-0.3cm}
\end{table}

\subsection{RQ3: How effective is TDL in generating intent tests?}
Since TDL is the core abstraction in our framework, it is essential to evaluate whether it faithfully captures test intent. 
Note that the impact of TDL on migration success is further examined in RQ4 through ablation studies. 
Here, we focus on two complementary perspectives: automated reconstruction fidelity and human validation. 

\subsubsection{Reconstruction study}
To evaluate whether TDL serves as a faithful abstraction of test intent, we conduct a reconstruction study. 
We randomly sample 200 tests from nine repositories and generate their corresponding TDL representations. 
Given the TDL and source repository context, an LLM is asked to reconstruct the original test.
We evaluate reconstruction fidelity by comparing the original test with the TDL-based reconstructed version using AST-based Jaccard Similarity. 
This metric reflects whether the TDL contains sufficient information to faithfully rebuild the original test, beyond surface-level variations such as identifier renaming.

\begin{figure}
\centering
\includegraphics[width=0.95\linewidth]{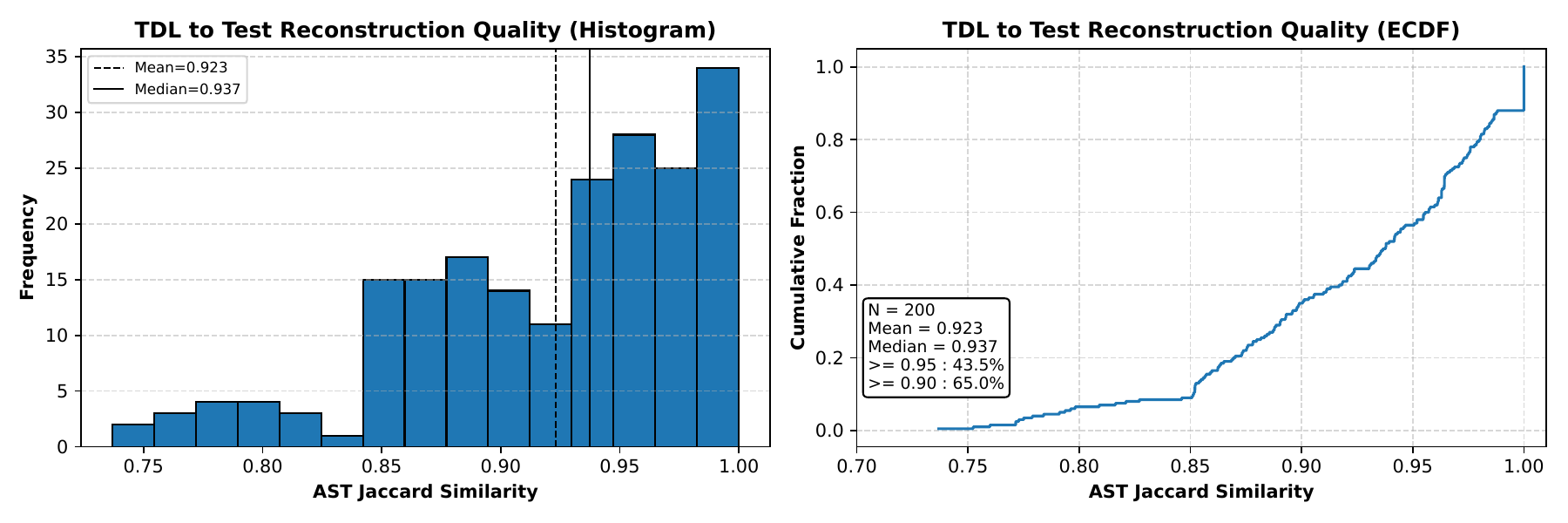}
\caption{Reconstruction quality from TDL: histogram (left) and ECDF (right) of AST Jaccard similarity, evidencing high fidelity; lower scores mainly reflect LLM refinements rather than semantic loss.}
\label{fig:tdl_reconstruction_quality}
\vspace{-0.3cm}
\end{figure}

As shown in Figure~\ref{fig:tdl_reconstruction_quality}, the reconstructed tests achieve consistently high similarity, with a mean of 0.92 and a median of 0.94. 
Nearly two-thirds of the cases exceed 0.90, and more than 40\% reach above 0.95, demonstrating that TDL preserves sufficient structural information to regenerate the original tests.
Further analysis reveals that cases with lower similarity do not indicate semantic loss, but rather LLM-driven refinements. 
For example, some reconstructed tests add descriptive messages to assertion statements, or replace \texttt{assertEqual} with semantically equivalent \texttt{assertTrue}. 
In exception handling, the model sometimes rewrites \texttt{assertThrows} as \texttt{@Test(expected=...)}, changing syntax but not behavior. 
These adjustments reduce structural similarity while maintaining the test’s functional intent.
TDL provides a compact and language-agnostic abstraction that enables faithful reconstruction of tests. 
Minor structural divergences reflect alternative, but valid, realizations of the same intent, underscoring TDL’s suitability as the foundation for cross-library test migration.

\begin{wrapfigure}{r}{0.5\textwidth}
\centering
\includegraphics[width=\linewidth]{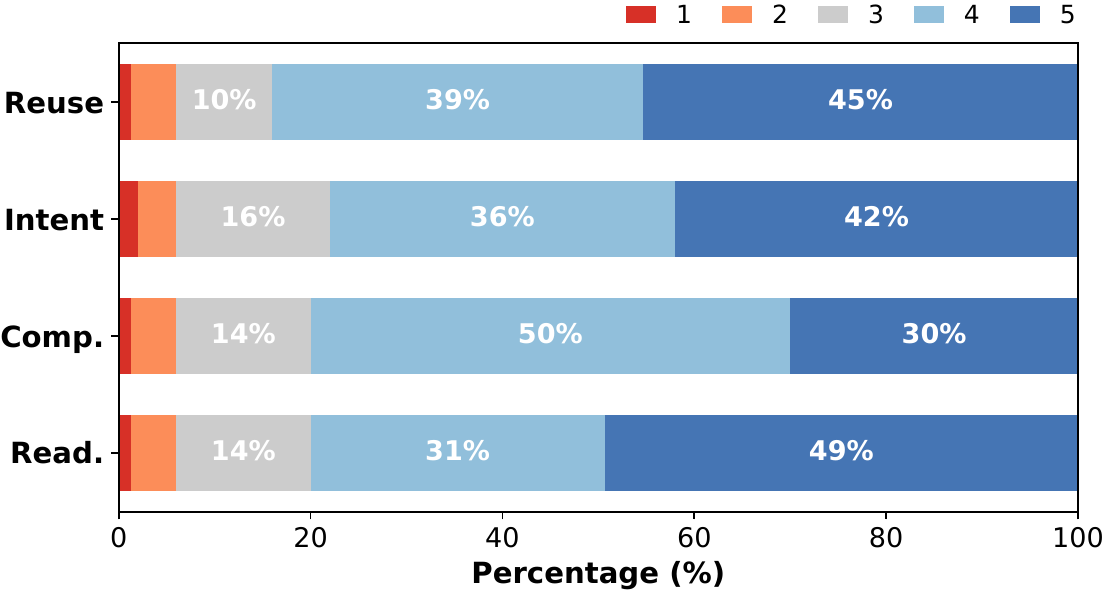}
\caption{Results of the user study on TDL.}
\label{fig:lierkt}
\end{wrapfigure}

\subsubsection{Human Validation}
To assess whether TDL faithfully captures test intent, we conduct a user study with three senior developers who each have over five years of experience in both Java and Python. 
From nine projects, we randomly select fifty TDL–test pairs and ask participants to rate TDL quality on four dimensions (Readability, Completeness, Intent, Reusability) on a 5-point Likert scale. 
Figure~\ref{fig:lierkt} shows the results: 80\% of readability scores are 4 or above, 80\% for completeness, 78\% for intent fidelity, and 84\% for reusability (with nearly half of the reusability scores rated 5). 
These outcomes suggest that TDL is generally clear, preserves the intent of the original test, and supports reuse across repositories and languages. 
Overall, TDL provides a reliable abstraction of test intent, capturing both the verification goal and sufficient structural information for cross-library reuse.

\textbf{Error Analysis}. We manually inspect the low-scoring cases (ratings of 3 or below) and identify two recurring issues. 
First, some descriptions are over-generalized, omitting key details such as exception types—for example, describing a step as \textit{verify error handling} without specifying the expected \texttt{IllegalArgumentException}, which reduces completeness. 
Second, some TDLs are overly verbose, using long phrases for simple operations (e.g., \textit{initialize an empty collection object and then proceed to append}) instead of concise ones (e.g., \textit{create an empty list}), which reduces readability. 
Both issues are minor, do not affect the correctness of subsequent test migration, and can be mitigated by refining the prompting strategy. 
Importantly, none of the developers report systematic misrepresentation of test intent, indicating that the errors concern expression quality rather than semantic accuracy.

\subsection{RQ4: Ablation Study.}

\begin{wraptable}{r}{0.5\textwidth}
\centering
\caption{Ablation study on 200 randomly sampled tests, CSR: Compilation Success Rate, EPR: Execution Pass Rate.}
\begin{tabular}{l|c|c}
\toprule
\textbf{Variant} & \textbf{CSR} & \textbf{EPR} \\
\midrule
w/o Intent Abstractor Agent       & 60\% & 54\% \\
w/o Relation-aware Expansion      & 76\% & 57\% \\
w/o Planning Agent                & 62\% & 49\% \\
w/o Verification Agent            & 68\% & 51\% \\
\textbf{\appname (Full) }         & \textbf{88\%} & \textbf{72\%} \\
\bottomrule
\end{tabular}
\label{tab:rq4_ablation}
\end{wraptable}

To assess the contribution of each agent, we conduct an ablation study on a random sample of 200 tests. 
We measure two metrics: (1) Compilation Success Rate, i.e., whether the migrated tests are syntactically valid and compilable, and 
(2) Execution Pass Rate, i.e., whether the tests execute successfully on the target repositories.
We compare \appname against four ablated variants:
\textit{w/o Intent Abstractor Agent (TDL)} – directly using raw source tests as queries, without intent-level abstraction.  
\textit{w/o Relation-aware Expansion} – performing only step-level semantic search, without dependency expansion.  
\textit{w/o Planning Agent} – skipping sufficiency validation and directly generating tests.  
\textit{w/o Verification Agent} – omitting post-generation checks, accepting tests without filtering.

As shown in Table~\ref{tab:rq4_ablation}, all components are critical. 
Removing TDL abstraction or Relation-aware Expansion leads to the largest drops (CSR ↓28\% and 12\%, EPR ↓18\% and 15\%), underscoring that intent-level representation and dependency modeling are key for retrieving executable contexts. 
Omitting Planning or Verification mainly affects robustness, with CSR falling to 62–68\% (vs. 88\%), since many generated tests contain unresolved symbols or inconsistent assertions that would otherwise be filtered. 
Overall, the agents complement each other: Intent Abstraction and Relation Expansion secure semantic and structural fidelity, while Planning and Verification ensure syntactic and execution-level reliability. 
Together, they enable \appname to reach the highest CSR (88\%) and EPR (72\%), confirming the necessity of the multi-agent design for robust, executable migration.

\section{Discussion}
\label{sec:discussion}
\subsection{Coverage Implications of Intent-Based Test Migration}
\begin{wrapfigure}{r}{0.55\textwidth}
\centering
\includegraphics[width=\linewidth]{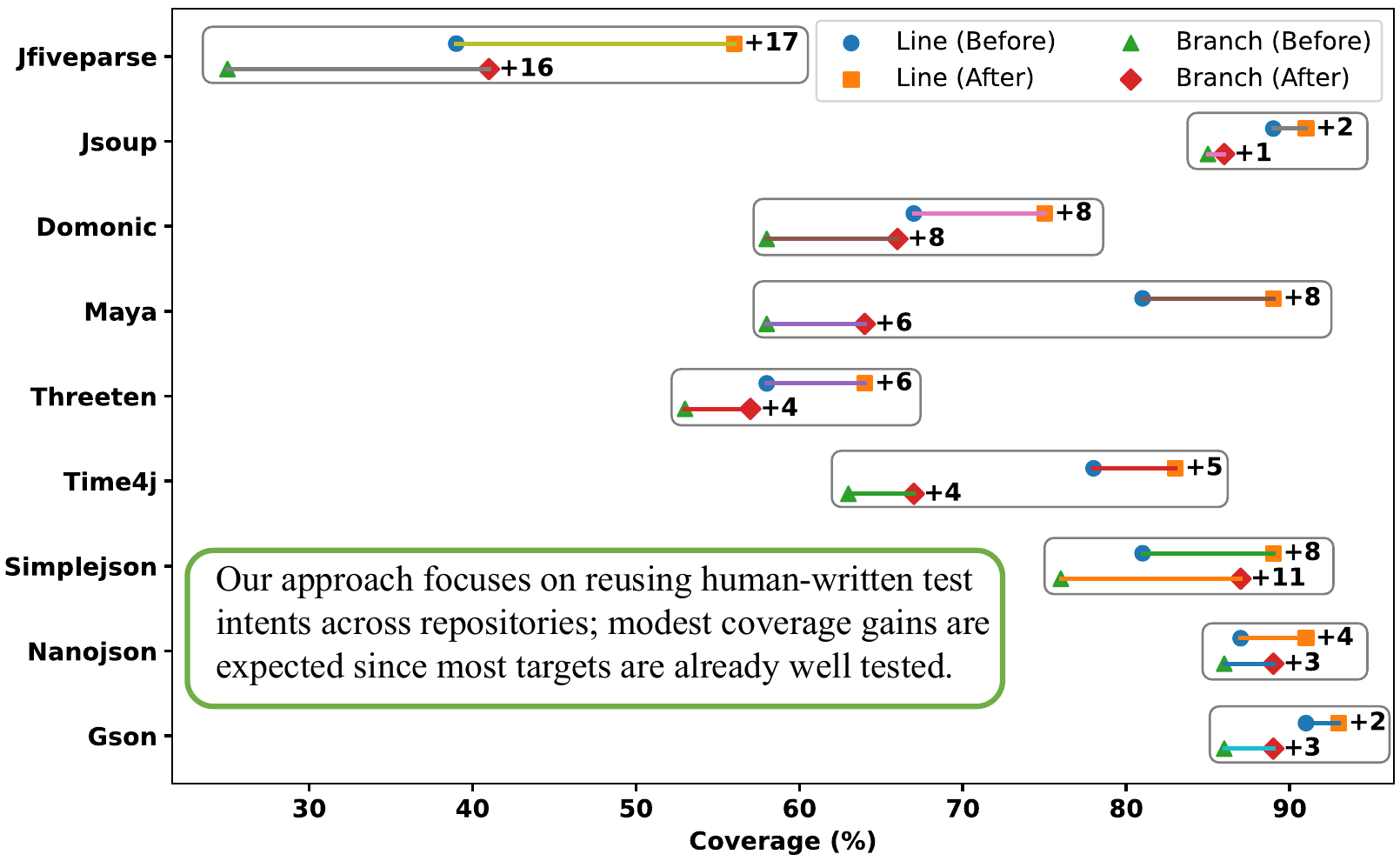}
\caption{Line and branch coverage achieved by \appname across target repositories, showing the additional coverage contributed by migrated intent tests.}
\label{fig:coverage}
\end{wrapfigure}

A natural question is whether intent-driven test migration with \appname improves code coverage.
We measured line and branch coverage before and after migration across nine repositories in three domains (Figure~\ref{fig:coverage}). 
The results show modest increases—for instance, 
\textit{JFiveParse} improves by +17\% in line coverage, and \textit{Domonic} by +8\%—while mature repositories such as \textit{Jsoup} and \textit{Gson} show only marginal gains (1–3\%).

This trend is expected given the motivation of \appname. 
Importantly, the design of \appname is not to maximize structural coverage, but rather to \textbf{reuse existing, human-written tests that encode domain knowledge across repositories}. 
Many source tests exercise core behaviors such as parsing, serialization, or argument validation—scenarios already well-covered in mature target repositories.
Consequently, migrating such tests rarely introduces additional coverage, but it still strengthens the test suite by (1) verifying semantic consistency across implementations, (2) uncovering previously untested failure modes (e.g., \texttt{NullPointerException} in \textit{JFiveParse}, \texttt{StackOverflowError} in \textit{NanoJSON}), and (3) enriching the diversity of assertion oracles.

Moreover, coverage gaps often arise from repository-specific features with no functional counterpart in the source repository. 
For example, \textit{JFiveParse} contains \texttt{TokenizerTagStates}, a specialized state-machine component for HTML tokenization not found in \textit{Jsoup} or \textit{Domonic}. 
Since no equivalent test intent exists, migrating tests cannot improve coverage in such regions.

Overall, the primary contribution of \appname lies in enabling cross-library test reuse and exposing hidden robustness issues, rather than simply boosting raw coverage metrics.
Future work could integrate intent reuse with coverage-guided generation to balance semantic alignment with structural exploration.

\subsection{Threats to Validity}
\textbf{Internal Validity.} A possible threat lies in the experimental setup, such as the choice of benchmarks and sampling strategy. 
We mitigate this by using nine diverse open-source libraries across three domains and two programming languages, and by reporting aggregated results over thousands of generated tests. 
Another internal threat is in the human evaluation of TDL fidelity, which could introduce subjectivity. 
To reduce bias, we employed multiple senior developers, used a structured Likert-scale rubric, and measured agreement across raters.
Finally, our implementation relies on a specific \textit{Llama-3.3-70B}. 
While the framework itself is model-agnostic, different LLMs may vary in their ability to capture domain semantics, which could affect absolute performance numbers. 
We mitigate this by reporting results across diverse repositories and ensuring that all baselines use the same model configuration for fairness.

\textbf{External Validity.} Our current evaluation is limited to Java and Python projects in three domains (JSON, HTML, and Time). 
While these domains are representative and cover both intra-language and cross-language migration, future work should explore additional ecosystems (e.g., C++, JavaScript) and industrial settings. 
Finally, while we observed real defect discovery and feature gaps, the scale of issue reporting remains limited; broader collaboration with maintainers would further validate practical utility.
\section{Related Work}
\label{sec:relatedwork}
\noindent \textbf{API Mapping and Test Migration.} 
Early work explored API mapping through call graph and structural analysis~\cite{teyton2013automatic,zhang2020deep,huang2024mapping,zhou2023hybrid,islam2024characterizing,shao2022cross,chen2026every}. 
MUT~\cite{gao2024mut} migrated unit tests by structural code mapping, while Nguyen et al.~\cite{nguyen2014statistical} applied statistical learning for cross-language method mapping. 
METALLICUS~\cite{sondhi2021mining} mined similar methods for automated test adaptation, and JTestMigrator~\cite{jha2024migrating} leveraged semantic similarity, achieving a 73\% success rate in Java ecosystems. 
However, these approaches primarily rely on syntactic translation, often failing to preserve original test intents when semantics diverge.

\noindent \textbf{Beyond GUI-Centric Migration.} 
Much prior work targeted GUI applications~\cite{behrang2018test,zhao2020fruiter,behrang2019test,hu2018appflow,lin2019test,qin2019testmig,mariani2021semantic,zhang2024synthesis,beyzaei2025automated,khalili2024semantic}. 
TestMig~\cite{qin2019testmig} translated cross-platform UI events, while TEMdroid~\cite{zhang2024learning} achieved 76\% widget-matching accuracy using BERT embeddings. 
Recent advances in GUI semantics~\cite{liu2022test,zhang2024learning,zhang2024llm,yoon2024intent} highlight progress in intent preservation at the UI level. 
In contrast, repository functionality tests across programming languages remain underexplored. 
Although MUT~\cite{gao2024mut} systematized unit test migration, it inherited the limitations of syntactic mapping. 
LLM-augmented techniques, such as AutoCodeRover~\cite{zhang2024autocoderover}, demonstrate promise in program repair through AST analysis, but naively applied LLMs often generate semantically inconsistent tests when bridging library gaps.

\section{Conclusion and Future Work}

This paper presents \appname, a multi-agent framework for intent-driven test migration that combines TDL abstraction, repository graphs, and LLM-guided synthesis.
By shifting from brittle structural mappings to semantic alignment of test intent, \appname enables automated cross-library and cross-language test migration without manual intervention.
Our evaluation on nine real-world projects shows that \appname achieves 85\% syntactic correctness and 74\% execution success, while also uncovering 25 real defects across widely used libraries. 
These results demonstrate both its practical effectiveness and its ability to strengthen test suites with minimal developer effort.
For future work, we plan to extend \appname to support additional programming languages and library ecosystems, and to investigate how different LLM sizes and configurations influence migration quality. 
This will further broaden its applicability and robustness in diverse software environments.
\section{Data Availability}
Our tool is available at~\cite{intenttest}.

\begin{acks}
This work was supported by the National Key R\&D Program of China (No. 2024YFB4506400).
\end{acks}

\bibliographystyle{ACM-Reference-Format}
\bibliography{main}
\end{document}